\documentclass[12pt,a4paper]{article}

\usepackage[dvips]{graphicx}

\begin{document}

\begin{center}{
{\bfseries {First result of the experimental search for the 9.4
keV solar axion reactions with $^{83}$Kr in the  copper
proportional counter}}\footnote{{\small Talk at The International
Workshop on Prospects of Particle Physics: "Neutrino Physics and
Astrophysics" January 26 - Ferbuary 2, 2014, Valday, Russia}}

\vskip 5mm

Yu.M. Gavrilyuk$^{a}$, A.M. Gangapshev$^{a}$, A.V. Derbin$^{b}$, V.V. Kazalov$^{a}$,\\
H.J. Kim$^c$, Y.D. Kim$^d$, V.V. Kobychev$^{e}$, V.V.
Kuzminov$^{a}$, Luqman Ali$^c$,\\ V.N. Muratova$^{b}$, S.I.
Panasenko$^{a,f}$, S.S. Ratkevich$^{a,f}$, D.A. Semenov$^b$,
\\ D.A. Tekueva$^{a}$, S.P. Yakimenko$^{a}$, E.V. Unzhakov$^b$
}
\end{center}

\begin{center}{

{\small {\it $^a$ Institute for Nuclear Research, RAS, Moscow,
Russia}}
\\
{\small {\it $^b$ Petersburg Nuclear Physics Institute, St.
Petersburg, Russia}}
\\
{\small {\it $^c$ Department of Physics, Kyungpook National
University, Daegu, Republic of Korea}}
\\
{\small {\it $^d$ Institute of Basic Science, Daejeon, Republic of
Korea}}
\\
{\small {\it $^e$  Institute for Nuclear Research of NAS Ukraine,
Kiev, Ukraine}}
\\
{\small {\it $^f$ Kharkov National University, Kharkov, Ukraine}}
}
\end{center}

\vskip 5mm

\centerline{\bf Abstract} {\small {The experimental search for
solar hadronic axions is started at the Baksan Neutrino
Observatory  of the Institute for Nuclear Researches Russian
Academy of Science (BNO INR RAS). It is assumed that axions are
created in the Sun during $M1$ transition between the first
thermally excited level at 9.4 keV and the ground state in
$^{83}\rm{Kr}$. The experiment is based on axion detection via
resonant absorption process by the same nucleus in the detector.
The big copper proportional counter filled with krypton is used to
detect signals from axions. The experimental setup is situated in
the deep underground low background laboratory. No evidence of
axion detection were found after the 26.5 days data collection.
Resulting new upper limit on axion mass is $m_{A} \leq$ 130 eV at
95\% C.L..}}
\\
\\
{\small \textbf{Keywords:} Solar axions; Axion-photon coupling; Axion-nucleon couplingr; Dark matter }

\section{Introduction}

A solution of the strong CP problem based on the global chiral symmetry U(1) was proposed by Peccei and Quinn (PQ) \cite{PQ}.  The existence of the
axions was predicted by Weinberg \cite{Wei} and Wilczek \cite{Wil} as result of spontaneous breaking of the PQ-symmetry at the energy $f_A$. The
axion mass ($m_A$) and the strengths of an axion's coupling to an electron ($g_{Ae}$), a photon ($g_{A\gamma}$ ) and nucleons ($g_{AN}$) are
proportional to the inverse of $f_A$. At the moment there are two classes of models for the axion: the hadronic axion (KSVZ-model) \cite{KSVZ} and
GUT axion (DFSZ-model) \cite{DFSZ}. The axion mass in both models is defined as:
\begin{eqnarray}\label{mass}
  m_A = \frac{f_\pi m_\pi}{f_A} \left(\frac{z}{(1+z+w)(1+z)}\right),
\end{eqnarray}
where $f_\pi \simeq 93$ MeV - pion decay constant, $z = m_u / m_d \simeq 0.56$ and $w = m_u / m_s \simeq 0.029$ - quark-mass ratios. It gives $ m_A
\left[ {\rm{eV}} \right] \simeq 6.0 \times 10^6 / f_A \left[ {\rm{GeV}} \right]$.

The main difference between models is that in contrast to the DFSZ-model in KSVZ-model axions have no coupling to leptons and ordinary quarks at the
tree level. As result the interaction of the KSVZ axion with electrons through radiatively induced coupling is strongly suppressed  \cite{SreKap}.


If axions do exist, then the Sun and other stars should be an intense source of these particles. In 1991 Haxton and Lee calculated the energy loss of
stars along the red-giant and horizontal branches due to the axion emission in nuclear magnetic transitions in $^{57}\rm{Fe}$, $^{55}\rm{Mn}$, and
$^{23}\rm{Na}$ nuclei \cite{HaxLee}. In 1995 Moriyama  proposed experimental scheme to search for 14.4 keV monochromatic solar axions that would be
produced when thermally excited $^{57}\rm{Fe}$ nuclei in the Sun relax to its ground state and could be detected via resonant excitation of the same
nuclide in a laboratory \cite{Mor95}. Searches for resonant absorption of solar axions emitted in the nuclear magnetic transitions were performed
with $^{57}\rm{Fe}$ \cite{Krc98,Der07,Nam07,Der09,Dan09,Der11}, $^{7}\rm{Li}$ \cite{Krc01,Der05,Bel12} and $^{83}\rm{Kr}$ \cite{Jak04} nuclei.

In this paper we present the results of the search for solar
axions using the the resonant absorption by $^{83}\rm{Kr}$ nuclei.
The energy of the first excited $7/2^+$ nuclear level  is equal to
9.405 keV, lifetime $\tau = 2.23\times10^{-7}$ s, internal
conversion coefficient $\alpha = 17.0$ and the mixing ratio of
$M1$ and $E2$ transitions is $\delta$ = 0.013 \cite{Wu01}.

In accordance with indirect estimates the abundance of the krypton in the Sun (Kr/H) = $1.78\times10^{-9}$ atom/atom \cite{Asp09} that corresponds to
$N= 9.08\times10^{13}$ of $^{83}\rm{Kr}$ atom per 1 g material in the Sun.

To estimate axion flux from the Sun, the calculation can be
performed as in the Ref.\cite{HaxLee}. The axion flux from a unit
mass is equal
\begin{eqnarray}\label{enlos}
  \delta \Phi(T) = N \frac{2 \exp(-\beta_T)}{1+2\exp(-\beta_T)} \frac{1}{\tau_\gamma} \frac{\omega_A}{\omega_\gamma},
\end{eqnarray}
where $N$ - number of $^{83}\rm{Kr}$ atoms in 1~g of material in
the Sun, $\beta_T=E_\gamma/kT$, $\tau_\gamma$ - lifetime of the
nuclear level, ${\omega_A}/{\omega_\gamma}$ - represents the
branching ratio of axions to photons emission. The ratio
$\omega_A/\omega_\gamma$ was calculated in
\cite{HaxLee,Don78,Avi88} as
\begin{eqnarray}\label{branch}
  \frac{\omega_{A}}{\omega_{\gamma}} =
\frac{1}{2\pi\alpha}\frac{1}{1+\delta^2}\left[\frac{g_{0}\beta+g_{3}}{(\mu_{0}-0.5)\beta+\mu_{3}-\eta}\right]^{2}
\left(\frac{p_{A}}{p_{\gamma}}\right)^{3},
\end{eqnarray}
where $\mu_0$ and $\mu_3$ - isoscalar and isovector magnetic moments, $g_{0}$ and $g_3$ - isoscalar and isovector parts of the axion-–nucleon
coupling constant $g_{AN}$ and $\beta$ and $\eta$ - nuclear structure dependent terms.

In case of the $^{83}\rm{Kr}$ nucleus, which has the odd number of nucleons and an unpaired neutron, in the one-particle approximation the values of
$\beta$ and $\eta$ can be estimated as $\beta\approx$-1.0 and $\eta\approx$0.5.

In the hadronic axion models, the  $g_0$ and $g_3$ constants can be represented in the form \cite{SreKap}:
\begin{eqnarray}\label{g0}
  g_{0}=-\frac{m_N}{6f_A}[2S+(3F-D)\frac{1+z-2w}{1+z+w}],
\end{eqnarray}
\begin{eqnarray}\label{g3}
 g_{3}=-\frac{m_N}{2f_A}[(D+F)\frac{1-z}{1+z+w}].
\end{eqnarray}
where $D$ and $F$ denote the reduced matrix elements for the SU(3) octet axial vector currents and $S$ characterizes
the flavor singlet coupling \cite{SreKap}. The parameter $S$ characterizing the flavor singlet coupling still remains a
poorly constrained one \cite{Der11}.  The most stringent boundaries $(0.37\leq S\leq0.53)$ and $(0.15\leq S\leq0.5)$
were found in \cite{Alt97} and \cite{Ada97}, accordingly.

The axion flux was calculated for the standard solar model BS05 \cite{Bah05} characterized by a highmetallicity \cite{Gre98}. The differential flux
at the maximum of the distribution (6) is
\begin{equation}\label{axionflux_num}
\Phi_{A}(E_{M1}) = 5.97\times 10^{23}\left(\frac{\omega_{A}}{\omega_{\gamma}}\right) \rm{cm}^{-2} \rm{s}^{-1}
\rm{keV}^{-1}.
\end{equation}
The width of the resulting distribution, which is described well by a Gaussian curve, is $\sigma$ = 1.2 eV. This value exceeds substantially the
recoil-nucleus energy and the intrinsic  and Doppler widths of the level of $^{83}\rm{Kr}$ target nuclei.

The cross section for resonance axion absorption is given by an expression similar to the expression for the
photon-absorption cross section, the correction for the ratio $\omega_A /\omega_{\gamma}$ being taken into account.
\begin{equation}\label{crosssection}
\sigma(E_{A})=2\sqrt{\pi}\sigma_{0\gamma}\exp\left[-\frac{4(E_{A}-E_{M})^{2}}{\Gamma^{2}}\right]\left(\frac{\omega_{A}}{\omega_{\gamma}}\right),
\end{equation}
where $\sigma_{0\gamma} = 1.22\times10^{-18} \rm{cm}^2$ is the
maximum cross section of the $\gamma$ -ray resonant absorption and
$ \Gamma= 1/\tau$. The total cross section for axion absorption
can be obtained by integrating $\sigma(E_A)$ over the axion
spectrum. The expected rate of resonance axion absorption by the
$^{83}\rm{Kr}$ nucleus as a function of the probability for axion
emission $\omega_A/\omega_{\gamma}$; the parameter $(g_{3} -
g_{0})$, which describes axion-nucleon interaction; and the axion
mass in the KSVZ model can be represented in the form ($S$ = 0.5,
$z$ = 0.56):
\begin{eqnarray}\label{count_speed}
R_A \rm{[g^{-1}day^{-1}]} = 4.23\times10^{21}(\omega_{A}/\omega_{\gamma})^2 \\ \label{count_speed_2} =
8.53\times10^{21}(g_3-g_0)^4(p_A/p_{\gamma})^6
\\ \label{count_speed_3}  = 2.41\times10^{-10}(m_{A})^{4}(p_A/p_{\gamma})^6.
\end{eqnarray}

The number of detected photons following axion absorption is determined by the target mass, measurement time, and
detector efficiency. At the same time, the probability of observing a peak at 9.4 keV is dependent on the background
level in the experimental facility used.

\section{Experimental setup}
The experimental technic is based on registration of the
$\gamma$-quantum and conversion electrons appearing after
deexcitation of the $^{83}$Kr nuclei. To register this process a
large proportional counter (LPC) with a casing of copper is used.
The gas mixture Kr(99.55$\%$)+Xe(0.45$\%$) is used as working
media of the LPC. The isotopic  composition of the krypton is:
$^{78}$Kr(0.002$\%$) + $^{80}$Kr(0.411$\%$) +
$^{82}$Kr(41.355$\%$) + $^{83}$Kr(58.229$\%$) +
$^{84}$Kr(0.003$\%$). The LPC is a cylinder with inner and outer
diameters of $137$ and $150$ mm, respectively. A gold-plated
tungsten wire of 10 $\mu$m in diameter is stretched along the LPC
axis and is used as an anode. To reduce the influence of the
counter edges on the operating characteristics of the counter, the
end segments of the wire are passed through the copper tubes (3 mm
in diameter and 38.5 mm in length) electrically connected to the
anode. These segments operate as an ionization chamber with no gas
amplification. Taking into account  teflon insulators dimensions,
the distance from operation region to the flange is 70 mm. The
fiducial length of the LPC is 595 mm, and the corresponding volume
is 8.77 $L$. Gas pressure is 5.6 bar, and corresponding mass of
the $^{83}$Kr-isotope in fiducial volume of the LPC is 101 g. The
LPC is surrounded by passive shield made of copper ($\sim$20 cm),
lead ($\sim$20 cm) and polyethylene (8 cm).  The setup is located
in the Deep Underground Low-Background Laboratory at BNO INR RAS
\cite{DULB}, at the depth of 4700 m w.e., where the cosmic ray
flux is reduced by $\sim 10^7$ times in comparison to that above
ground, and evaluated as $(3.0 \pm 0.1) \times
10^{-9}$~cm$^{-2}$s$^{-1}$ \cite{Gav}.

\begin{figure}[pt]
\begin{center}
\includegraphics*[width=2.75in,angle=270.]{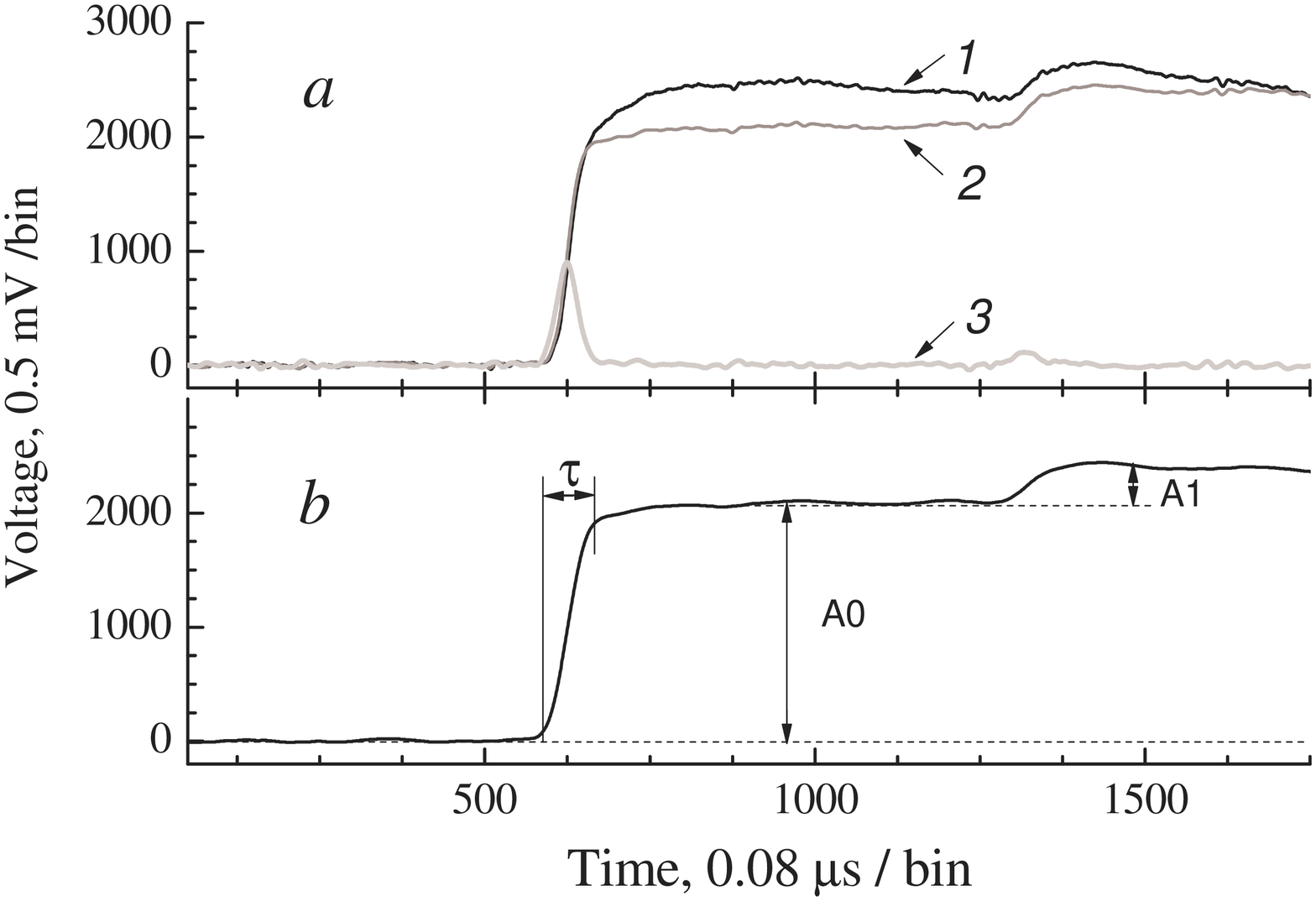} \caption{\label{fig:pulsam} \emph{a}) Samples of recorded pulse
(\emph{1}), electronic current (\emph{3}) and corrected pulse
(\emph{2}); \emph{b}) chematic description of the parameters of
the corrected pulses used for data analysis. A0 - amplitude of the
primary pulse, A1 - amplitude of the secondary pulse, $\tau$ -
pulse rise time.}
\end{center}
\end{figure}

The detector signals are passed from one end of the anode wire to
the charge-sensitive preamplifier. A total shape of  pulses
(waveforms) are recorded by the digitizer after amplification in
an auxiliary amplifier. A digital oscilloscope LAn-n20-12USB
integrated with a computer via USB-port is used. Sampling
frequency is 12.5 MHz and record window is 2048 bins (164 $\mu$s).
Sample of the recorded waveform is presented in Fig.\ref
{fig:pulsam}\emph{a}. There is a feature in the waveforms, namely
the presence of secondary pulse after the primary pulse. Those
secondary pulses are result of photoemission from the cathode.
Photons appear near anode wire in time of the development of an
electron avalanche from the primary ionization electrons. The
relative height of the secondary pulse depends on the position of
the events along anode wire due to the changing of a cathode view
solid angle. The lowest value of the angle is reached at the ends
of the anode wire and the smallest secondary pulses are appeared
as a result. Recorded waveforms are processed to define parameters
of the events. First of all, the response function of the LPC for
single electron of the primary ionization is used to define the
electron density profile for the given event (electron current).
The shape of the responce function is defined as:
\begin{eqnarray}\label{pulmod}
V(t)=a \cdot \exp^{ -\left (\frac{t+B1}{\tau} \right ) } \left (
\log\left ( 1+\frac{t}{B}\right ) +\frac{t}{\tau} +
\frac{t^{2}}{8\tau^{2}} \right ) \cdot \cos \left (\frac{2\pi
t}{T}\right ),
\end{eqnarray}
where $T=460$ $\mu$s, $B=2.4\cdot 10^{-3}$~$\mu$s and $\tau=240$~$\mu$s. The cumulative sum (integral) of the
electronic  current is an corrected waveform as in the Refs. \cite{PTE}. Samples of the recorded electronic current and
the corrected waveform are presented in Fig.\ref{fig:pulsam}\emph{a}. Schematic description of the used parameters are
presented in Fig.\ref{fig:pulsam}\emph{b}. The amplitude of the secondary pulse is used to define the relative position
of the events along the anod wire by ratio:
\begin{eqnarray}\label{lambda}
\lambda=1000 \cdot \frac{A1}{A0}
\end{eqnarray}
where $A1$ - amplitude of the secondary pulse, $A0$ - amplitude of the primary pulse (see Fig.\ref{fig:pulsam}\emph{b}).

\section{Results}

\begin{figure}[pt]
\begin{center}
\includegraphics*[width=2.75in,angle=0.]{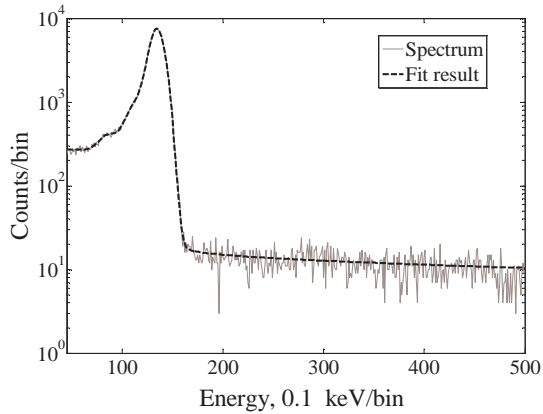}
\caption{\label{fig:spectra} Background spectra colected with LPC
during 26.5 days of measurements and fitting function
(\ref{fit}).}
\end{center}
\end{figure}

\begin{figure}[pt]
\begin{center}
\includegraphics*[width=2.75in,angle=0.]{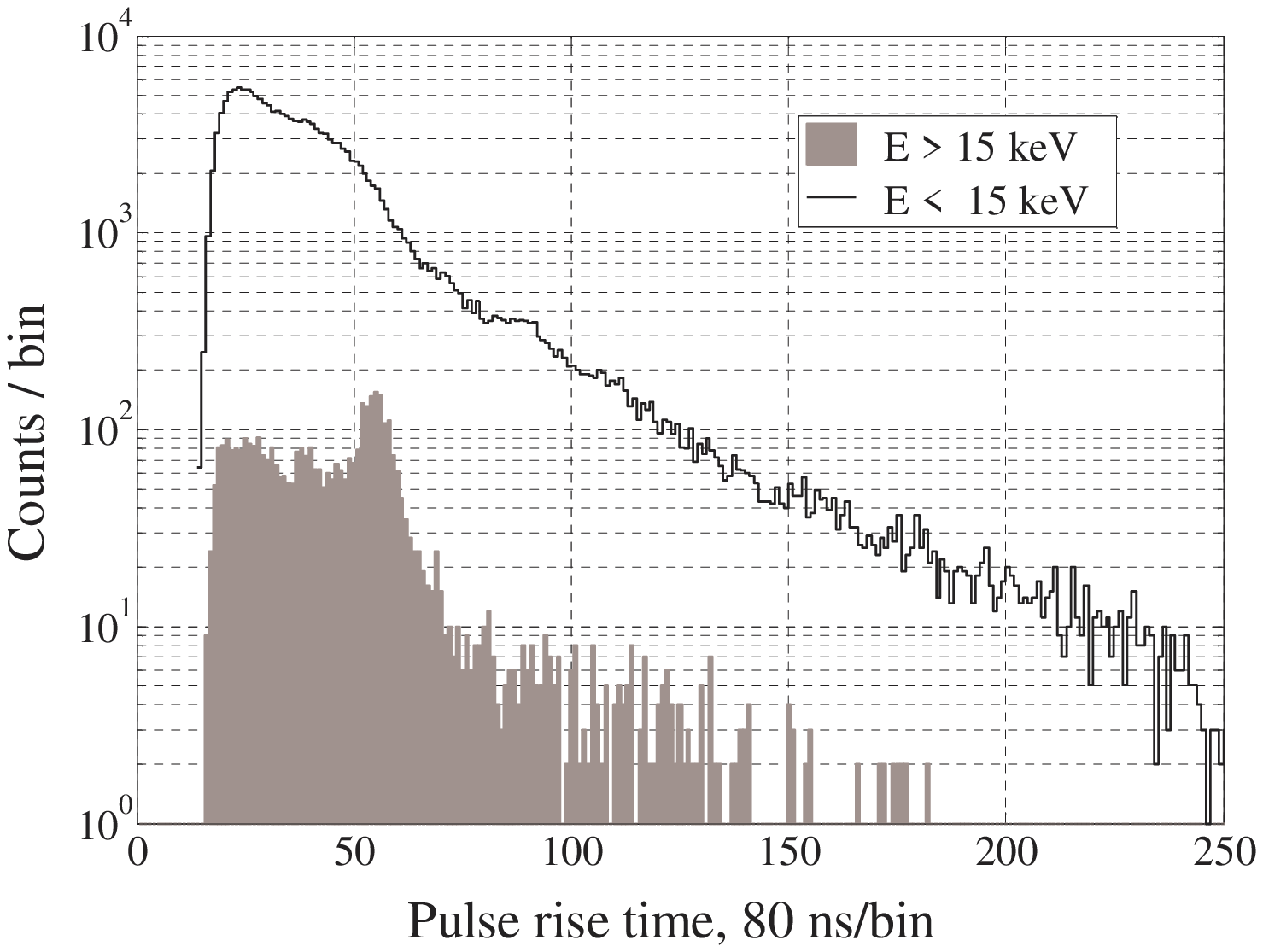}
\caption{\label{fig:risetime} Distribution of the events versus
pulse rise time.}
\includegraphics*[width=2.75in,angle=0.]{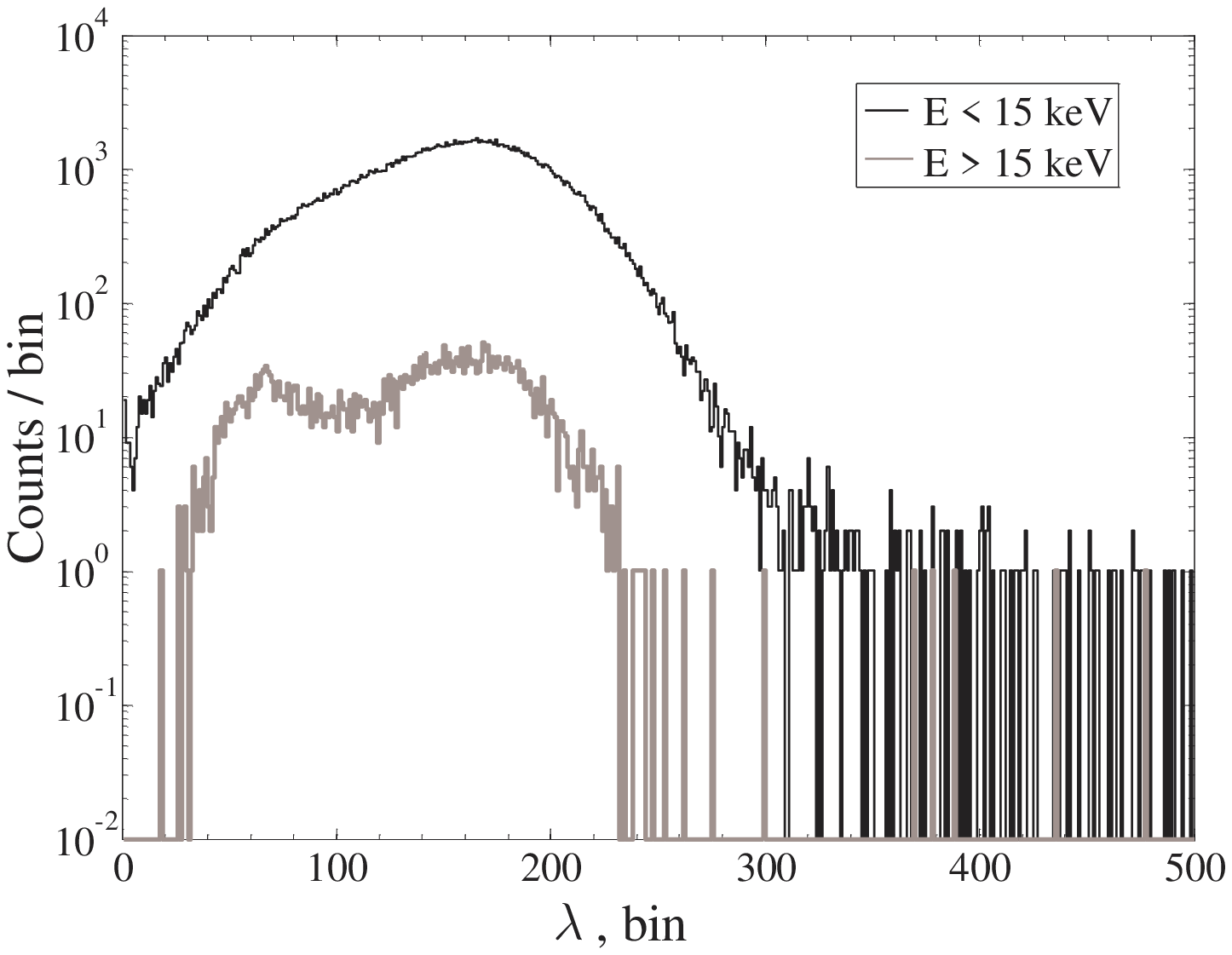}
\caption{\label{fig:lambda} Distribution of the events versus
$\lambda$.}
\end{center}
\end{figure}

The background spectra collected during 26.5 days and fit result curve are presented in Fig.\ref{fig:spectra}.
Fitting function is taken as
\begin{eqnarray}\label{fit}
f(x)=7151 \cdot e^{-\left ( \frac{x-134.6}{9.567} \right )^2}+886.4 \cdot e^{-\left ( \frac{x-119.1}{16.64} \right )^2}+130.7 \cdot e^{-\left ( \frac{x-85.72}{12.00} \right )^2} + \nonumber\\
+81.40 \cdot \left (1.57-\arctan \left ( \frac{x-123.4}{1.716}
\right )  \right ) + 10.88 \cdot e^{- \left ( \frac{x}{569.6}
\right )} + 5.596.
\end{eqnarray}
The peak of 13.5 keV from $K$-capture of $^{81}$Kr is well seen.
$^{81}$Kr is a cosmogenic isotope. It present as a trace admixture
in the used samples of krypton. This isotope is produced in
atmosphere mainly in reactions $^{82}$Kr($n,2n$)$^{81}$Kr and
$^{80}$Kr($n,\gamma$)$^{81}$Kr, the half-life time is
$T_{1/2}=2.1\cdot10^{5}$~y. There are two more peaks: $\sim 12$
and $\sim 8$ keV. The peak near 12 keV is produced by events,
where the $K$-capture of $^{81}$Kr happens near the end of the
counter (out of the fiducial volume) and the characteristic photon
of daughter nuclide $^{81}$Br is registered in the fiducial volume
($K_{\alpha1}$=11.923 keV (57.1\%), $K_{\alpha2}$=11.877 keV
(29.1\%), $K_{\beta1}$=13.290 keV (12.7\%),
$K_{\beta2}$=13.465~keV (1.0~\%) \cite{Blokhin}). The peak near
8~keV is produced by characteristic photons from the copper
($K_{\alpha1}$=8.047~keV (58.4~\%), $K_{\alpha2}$=8.027~keV
(29.8\%), $K_{\beta1}$=8.904~keV  (11.7\%) \cite{Blokhin}). The
distributions of the events versus pulse rise time and parameter
$\lambda$ are presented in Fig.\ref{fig:risetime} and
Fig.\ref{fig:lambda}
respectively. The events with pulse rise time $\simeq55$ bins
($4.4$~$\mu$s) mostly are the single site events from the inner
surface of the cathode (well seen in the distribution for the
events with $E>15.0$~keV). The pulses with londer pulse rise time
are mostly the multi site events. The events with $\lambda < 115$
are mostly close to the edge of the fiducial volume (events in the
peak of $\simeq 12$ keV) or out of it (high energy events at the
ends of the counter, where no gas amplification).

\begin{figure}[pt]
\begin{center}
\includegraphics*[width=2.75in,angle=0.]{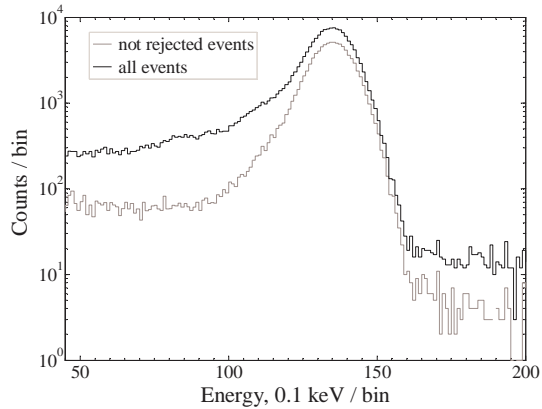}
\caption{\label{fig:compare} Original energy spectrum and spectrum
after rejection of the events with pulse rise time longer then 47
bins (3.8 $\mu$s) and  $\lambda$ lower then 115.}
\end{center}
\end{figure}

Thus, as we are looking for single site events in the inner volume of the detector, the events with pulse rise time longer  then 47 bins (3.8~$\mu$s)
and  $\lambda$ lower then 115 are rejected. The resulting spectrum in comparison with original one is presented in Fig.\ref{fig:compare}.
There is no visible peak around 9.4 keV from axions. Region of
interest is $8.2 \div 9.4$ keV, where should be 47.5\% of axion
events. Detection efficiencies after applying energy cut, pulse
rise time cut and $\lambda$ cut are $\varepsilon_E=0.475$,
$\varepsilon_\tau=0.94$, $\varepsilon_\lambda=0.91$ respectively.
The number of events in the region of interest after cuts applying
equal to 703 cts. As there is no "signal" from axions, the upper
limit on excitation rate of $^{83}$Kr by solar hadronic axions in
the detector is defined as:
\begin{eqnarray}\label{ratelim}
R_{exp} \leq \frac{2\sqrt{2B}}{m_{\rm{Kr}}~t_{\rm{meas}}~{\varepsilon_E}~{\varepsilon_{\tau}}~{\varepsilon_{\lambda}}} = 0.069~~
[\rm{g}^{-1}\rm{day}^{-1}],
\end{eqnarray}
where: $B$ - background in the region of interest, $m_{\rm{Kr}}$ - the mass of the $^{83}$Kr (gram), $t_{\rm{meas}}$ -
measurement time (day).

The relation (\ref{ratelim}) obtained in the experiment limits the region of possible values of the coupling constants $g_0$, $g_3$ and axion mass
$m_A$. In accordance with Eqs. (\ref{count_speed}-\ref{count_speed_3}), and on condition that $(p_A/p_\gamma)\cong 1$ provided for $m_A < 3$ keV one
can obtain:
\begin{equation}\label{limgAN}
|g_3-g_0|\leq 1.69\times 10^{-6}, \rm{~and}
\end{equation}
\begin{equation}\label{limma}
m_A \leq 130 \rm{~eV~~ at~ 95\%~ C.L.}
\end{equation}

The limit (\ref{limma}) is stronger then the constrain obtained with 14.4 keV  $^{57}\rm{Fe}$ solar axions - ($m_A\leq$ 145 eV \cite{Der11}) and is
significantly  stronger then previous result obtained in $^{83}$Kr experiment \cite{Jak04}.


\section{Conclusion.}
A search for resonant absorption of the solar axion by $^{83}\rm{Kr}$ nuclei was performed using the proportional counter installed inside the
low-background setup at the Baksan Neutrino Observatory.  The intensity of the 9.4 keV peak measured for 26.5 days turned out to be $\leq 7.0$
events/day. The obtained model independent upper limit on axion-nucleon couplings allowed us to set the new upper limit on the hadronic axion mass of
$m_{A}\leq 130$ eV (95\% C.L.) with the generally accepted values $S$=0.5 and $z$=0.56. The obtained limit on axion mass strongly depends on the
exact values of the parameters $S$ and $z$.

The work is supported by Russian Foundation of Basic Research (Grants No. 14-02-00258A, 13-02-01199A and 13-02-12140-ofi-m). V.V.Kobychev is
supported in part by the Space Research Program of NAS Ukraine.


\begin{thebibliography}{99}

\bibitem{PQ}
            \textit{Peccei R. D. and Helen R. Quinn} //
            {Phys. Rev.} {D. 1977. V.16.} P.1791;
\bibitem{Wei}
            \textit{Weinberg S.} //
            {Phys. Rev. Lett.} {1978. V.40.} P.223;
\bibitem{Wil}
            \textit{Wilczek F.} //
            {Phys. Rev. Lett.} {1978. V.40.} P.279;
\bibitem{KSVZ}
            \textit{Kim J.E.} //
            {Phys. Rev. Lett.} {1979. V.43.} P.103;\\
            \textit{Shifman M.A., Vainstein A.I., Zakharov V.I.} //
            {Nucl. Phys.} {B. 1980. V.166.} P.493;
\bibitem{DFSZ}
            \textit{Zhitnitskii A.R.} //
            {Yad. Fiz.} {1980. V.31.} P.497\\
            \textit{Dine M., Fischler F., Srednicki M.} //
            {Phys. Lett.} {B. 1981. V.104.} P.199;
\bibitem{SreKap}
            \textit{Srednicki M.} //
            {Nucl. Phys.} {B. 1985. V.260.} P.689;\\
            \textit{Kaplan David B.} //
            {Nucl. Phys.} {B. 1985. V.260.} P.215;
\bibitem{HaxLee}
            \textit{Haxton W.C. and Lee K.Y.} //
            {Phys. Rev. Lett.} {1991. V.66.} P.2557;

\bibitem{Mor95}\textit{Moriyama S.} // Phys. Rev. Lett. 1995. V.75 P.3222;
\bibitem{Krc98}\textit{Krcmar M. et al.} // Phys. Lett. B 1998. V.442 p.38;
\bibitem{Der07}\textit{Derbin A.V. et al.} // JETP Lett. 2007. V.85 p.12;
\bibitem{Nam07}\textit{Namba T.} // Phys. Lett. B 2007. V.645 p.398;
\bibitem{Der09}\textit{Derbin A.V. et al.} //{Eur. Phys. J.} {C 2009. V.62.} P.755;
\bibitem{Dan09}\textit{Danevich F.A. et al.} // {Kinematics and Physics of Celestial Bodies.} 2009. V25. P.102;
\bibitem{Der11}\textit{Derbin A.V. et al.} //Phys. At. Nucl. 2011. V.74. p.596;

\bibitem{Krc01}\textit{Krcmar M. et al.} // Phys. Rev. D 2001. V.64 p.115016;
\bibitem{Der05}\textit{Derbin A.V. et al.} //JETP Lett. 2005. V.81. p.365;
\bibitem{Bel12}\textit{Belli P. et al.} // Phys. Lett. B 2012. V.711. P.41;

\bibitem{Jak04}\textit{Jakov$\check{c}$i\'c K.  et al.} //{Radiat.Phys.Chem.} 2004. V71. P.93; arXiv:nucl-ex/0402016v1;

\bibitem{Wu01} \textit{Wu S.C.} // Nuclear Data Sheets 2001. V.92 p.893;
\bibitem{Asp09}\textit{Asplund M., Grevesse N., Sauval A.J., Scott P.} // Ann. Rev. of Astronomy and Astrophysics.
 2009. V.47 P.481; arXiv:0909.0948
\bibitem{Don78}\textit{Donnelly  T.W. et al.} // Phys. Rev. D 1978. V.18  P.1607;
\bibitem{Avi88}\textit{Avignone III F.T. et al.} // Phys. Rev. D 1988. V.37 p.618;
\bibitem{Alt97}\textit{Altarelli G., et al.} // Phys.~Lett. B 1997. V.46 p.337;
\bibitem{Ada97}\textit{Adams D. et al.} // Phys.~Rev. D 1997. V.56 p.5330;
\bibitem{Bah05}\textit{Bahcall J.H., Serenelli A.M., Basu S.} Astrophys. J. 2005. V.621 p.L85;
\bibitem{Gre98}\textit{Grevesse N., Sauval A.J.} Space Sci. Rev. 1998. V.85 p.161;


\bibitem{DULB}
            \textit{Gavriljuk Ju.M. et al.} //
            {Nucl. Ins. Meth.} {A. 2013. V.729.} P.576;

\bibitem{Gav}
            \textit{Gavrin V.N. et al.} //
            {Preprint INR RAS},{1991. P-698};

\bibitem{PTE} \textit{Gavrilyuk~Yu.M.  et al.}, // {Instr.Exper.Techn.} 2010. V53. P.57;
            \textit{Gavrilyuk~Yu.M.  et al.} // {Phys. Rev. C} 2013. V.87. P.035501.

\bibitem{Blokhin}
            \textit{Blokhin M.A., Shveizer I.G.}
            {Rentgenospektralniy spravochnik.}
            {Moskva, "Nauka", 1982.} (in russian)


\end{thebibliography}
\end{document}